\documentclass[sigconf]{acmart}

\setcopyright{none}
\renewcommand\footnotetextcopyrightpermission[1]{}
\usepackage{amsmath}
\usepackage{booktabs}
\usepackage{multirow}
\usepackage{tcolorbox}
\usepackage{dblfloatfix}
\usepackage{placeins}
\usepackage{url}
\usepackage{xcolor}
\usepackage{graphicx}
\usepackage{makecell}
\usepackage[table]{xcolor}
\usepackage{colortbl}
\usepackage{array}
\usepackage{microtype}
\usepackage{enumitem}

\usepackage{xcolor}
\newcommand{\gain}[1]{\textcolor{green!60!black}{\scriptsize$\uparrow$#1}}
\newcommand{\loss}[1]{\textcolor{red!70!black}{\scriptsize$\downarrow$#1}}

\title{Financial Transaction Retrieval and Contextual Evidence for Knowledge-Grounded Reasoning}

\author{Artem Sakhno}
\author{Daniil Tomilov}
\affiliation{
  \institution{Sber AI Lab}
  \city{Moscow}
  \country{Russian Federation}}

\author{Yuliana Shakhvalieva}
\author{Inessa Fedorova}
\author{Daria Ruzanova}
\affiliation{
  \institution{Sber AI}
  \city{Moscow}
  \country{Russian Federation}}

\author{Omar Zoloev}
\author{Andrey Savchenko}
\author{Maksim Makarenko}
\affiliation{
  \institution{Sber AI Lab}
  \city{Moscow}
  \country{Russian Federation}}

\renewcommand{\shortauthors}{Artem Sakhno et al.}
\ccsdesc[500]{Information systems~Information retrieval}
\ccsdesc[300]{Information systems~Data mining}

\usepackage{graphicx}

\newcommand{\methodname}{\textsc{FinTRACE}}

\newcommand{\maxim}[1]{{\color{blue} #1}}

\definecolor{bestgold}{HTML}{F6E58D}
\definecolor{secondsilver}{HTML}{DCE3EA}
\definecolor{thirdbronze}{HTML}{E8C3A4}

\newcommand{\best}[1]{\cellcolor{bestgold}\textbf{#1}}
\newcommand{\second}[1]{\cellcolor{secondsilver}\textbf{#1}}
\newcommand{\third}[1]{\cellcolor{thirdbronze}\textbf{#1}}

\setlist{topsep=2pt,itemsep=2pt,parsep=0pt,partopsep=0pt}

\renewcommand{\arraystretch}{1.18}
\keywords{financial transactions, retrieval-augmented reasoning, behavioral modeling, LLM grounding, structured data}

\begin{document}

\begin{abstract}
Nowadays, success of financial organizations heavily depends on their ability to process  digital traces generated by their clients, e.g., transaction histories, gathered from various sources to improve user modeling pipelines. 
As general-purpose LLMs struggle with time-distributed tabular data, production stacks still depend on specialized tabular and sequence models with limited transferability and need for labeled data. To address this, we introduce FinTRACE, a retrieval-first architecture that converts raw transactions into reusable feature representations, applies rule-based detectors, and stores the resulting signals in a behavioral knowledge base with graded associations to the objectives of downstream tasks. Across public and industrial benchmarks, FinTRACE substantially improves low-supervision transaction analytics, doubling zero-shot MCC on churn prediction performance from 0.19 to 0.38 and improving 16-shot MCC from 0.25 to 0.40. We further use FinTRACE to ground LLMs via instruction tuning on retrieved behavioral patterns, achieving state-of-the-art LLM results on transaction analytics problems.
\end{abstract}

\maketitle
% --------------------------------------------------

% --------------------------------------------------
\section{Introduction}

Production financial platforms typically rely on processing client's digital traces using collections of task-specific pipelines built from feature-engineered tabular models~\cite{faubel2024mlops, louhi2023empirical}, sequence learners~\cite{latte, shestov2025llm4es}, and rule-based systems~\cite{arms}, each optimized for a single downstream objective such as risk scoring~\cite{alphabattle}, churn detection~\cite{rosbank1}, or compliance monitoring~\cite{compliance}. While effective in isolation, these pipelines fragment evidence across systems and require repeated data preparation, retraining, and validation whenever a new use case emerges. As a result, behavioral data cannot be easily reused across tasks~\cite{cotic, featurestore}, limiting the ability to expose transactional data traces through a shared interface for flexible downstream reasoning~\cite{toolllm, react}.

\begin{figure}[!t]
  \centering
  \includegraphics[width=1\columnwidth]{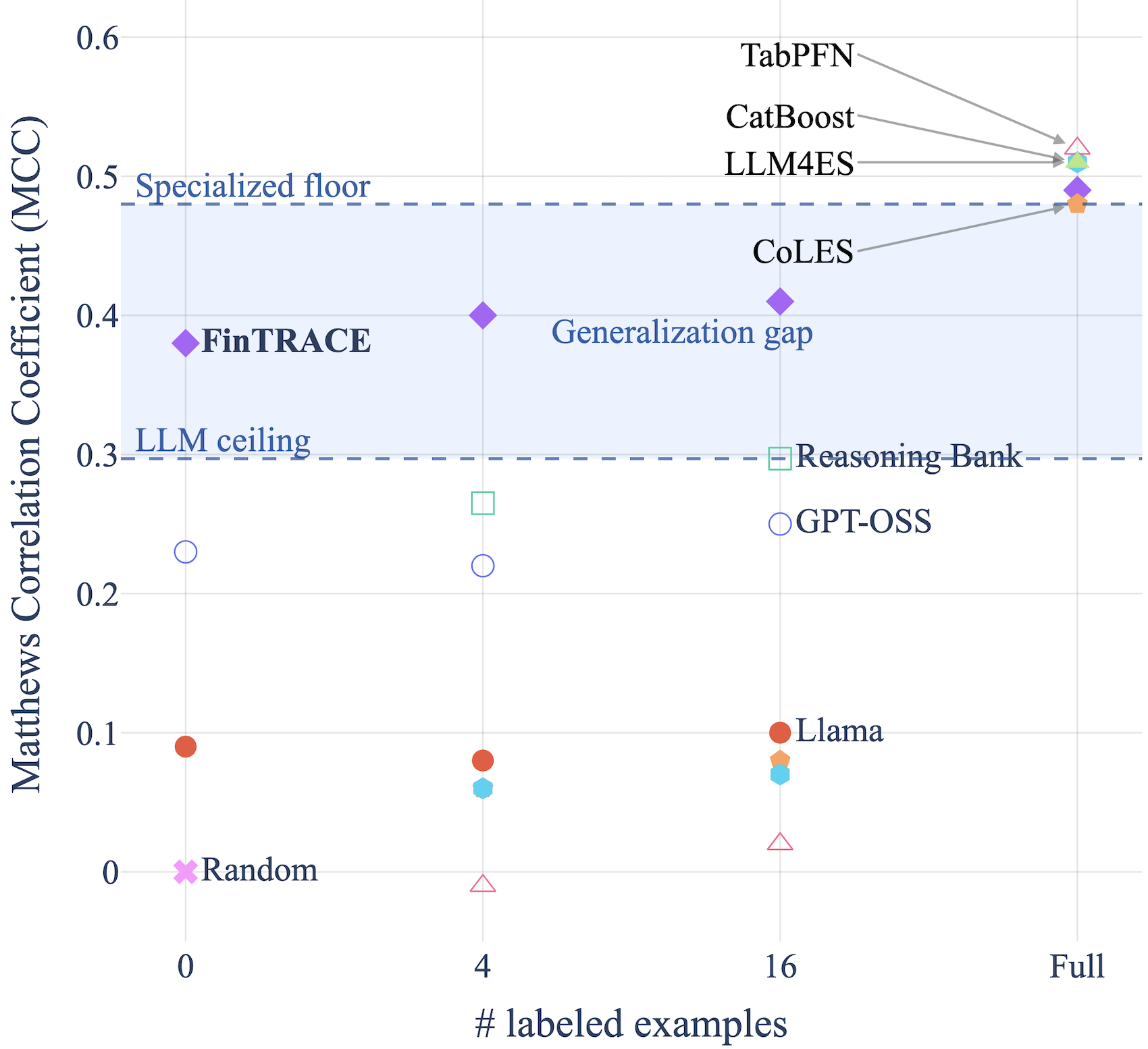}
% Shortened ~2
\caption{
\textbf{Generalization gap between LLM and specialized pipelines.}
In few-shot settings (0/4/16 labels), LLM methods (GPT-OSS, Llama-3-Instruct, Reasoning Bank) plateau below 0.30 MCC, while specialized models (TabPFN, CatBoost, LLM4ES, CoLES) exceed 0.48 MCC only with full supervision. The shaded region marks this gap. This work (\methodname) narrows it, achieving 0.38 MCC in zero-shot and 0.48 with full labels.
}
  \label{fig:label-efficiency}
\end{figure}

Large Language Models (LLMs) are natural candidates for such a unified interface, but applying them directly to transactional data remains difficult. 
Transaction histories are long irregular event sequences~\cite{ebes, localglobal} whose meaning is distributed across heterogeneous attributes (merchant identifiers, category codes, timestamps, amounts) that follow institution-specific conventions and often lack stable linguistic semantics. 
Naively serializing rows into prompts produces readable text, but semantically brittle evidence for downstream reasoning~\cite{radar, torr}.

Figure~\ref{fig:label-efficiency} provides direct evidence of this bottleneck. Off-the-shelf LLMs (GPT-4o, GPT-OSS~\cite{openai2025gptoss120bgptoss20bmodel}) remain in a low-performance regime under 0/4/16-shot supervision, indicating that transaction behavior is not natively represented in language space. 
Specialized tabular models (TabPFN~\cite{tabpfn}, CatBoost~\cite{catboost}) achieve much higher scores only in the fully labeled setting, suggesting strong label dependence. The shaded band between the LLM ceiling and specialized floor visualizes a persistent generalization gap between language-native and task-specific pipelines.

Existing efforts to bridge LLMs and structured data, such as prompt serialization for tabular reasoning (e.g., TabLLM~\cite{hegselmann2023tabllm}, Chain-of-Table~\cite{wangchain}), instruction tuning on large table corpora (e.g., TableLlama~\cite{zhang2024tablellama}, StructLM~\cite{zhuangstructlm}, TableGPT~\cite{li2024table,yang2025tablegpt}), and retrieval combined with domain-adapted models such as BloombergGPT~\cite{wu2023bloomberggpt} and FinGPT~\cite{yang2023fingpt}, largely assume that structured inputs can be rendered into language with limited loss of meaning.  %\maxim{please add citations here ASAP} 
Transactional data is more challenging~\cite{latte,shestov2025llm4es}: its predictive signal is distributed across temporal regularities, amount dynamics, and schema-specific attribute interactions rather than explicit textual semantics. As a result, the main bottleneck is not text generation itself, but the lack of a suitable knowledge-grounding~\cite{ograg,pathgraph} to organize behavioral evidence into a structured form for retrieval and reasoning. %While ontology- and KG-grounded methods have recently shown promise for improving factual grounding and structured reasoning in adjacent domains~\cite{ograg,pathgraph}, such approaches remain underexplored for behavioral transaction modeling.

We address this limitation with \methodname\ (Financial Transaction Retrieval and Context Engine)\footnote{\url{https://github.com/warofgam/FinTRACE}}, a retrieval-first framework that represents transactional data as a structured knowledge base of grounded behavioral evidence and rules. Rather than converting raw event logs directly into text, \methodname\ organizes transaction histories into reusable feature essences, higher-level behavioral patterns, and their relations to downstream targets, using lightweight white-box rules and statistical detectors. In this sense, the proposed approach is closer to ontology-based~\cite{retrievalaugmentedgeneration} and neuro-symbolic reasoning systems~\cite{bhuyan2024neuro,knowledgegraphreasoning}: it builds an explicit intermediate layer of behavioral evidence that can be retrieved, composed, and reused across objectives. Empirically, \methodname\ achieves new state-of-the-art results on both open-source and industrial benchmarks in low-supervision regimes. In addition, instruction-tuned LLMs grounded with \methodname\ reach performance (0.48 MCC) competitive with specialized transactional models such as CoLES (0.48 MCC) and LLM4ES (0.51 MCC), while preserving general text capabilities.

We summarize here the main contributions: (i) we introduce a behavioral indexing framework that decomposes raw transaction sequences into a reusable financial Knowledge Base (KB) and showcase how the KB can be used to efficiently form a retrieval context for LLMs; (ii) we demonstrate that the same KB can be converted into an instruction dataset for supervised instruction tuning, enabling domain adaptation without manual annotation; and (iii) we study a transfer setting in which the target is not used during KB construction and propose an inference-time self-reflection pipeline that recombines behavioral patterns learned from synthetic objectives to generalize to unseen financial tasks in zero- and few-shot regimes.

% --------------------------------------------------
\section{Proposed Methodology}

\label{sec:method}

\begin{figure}[h]
    \centering
    \includegraphics[width=1\linewidth]{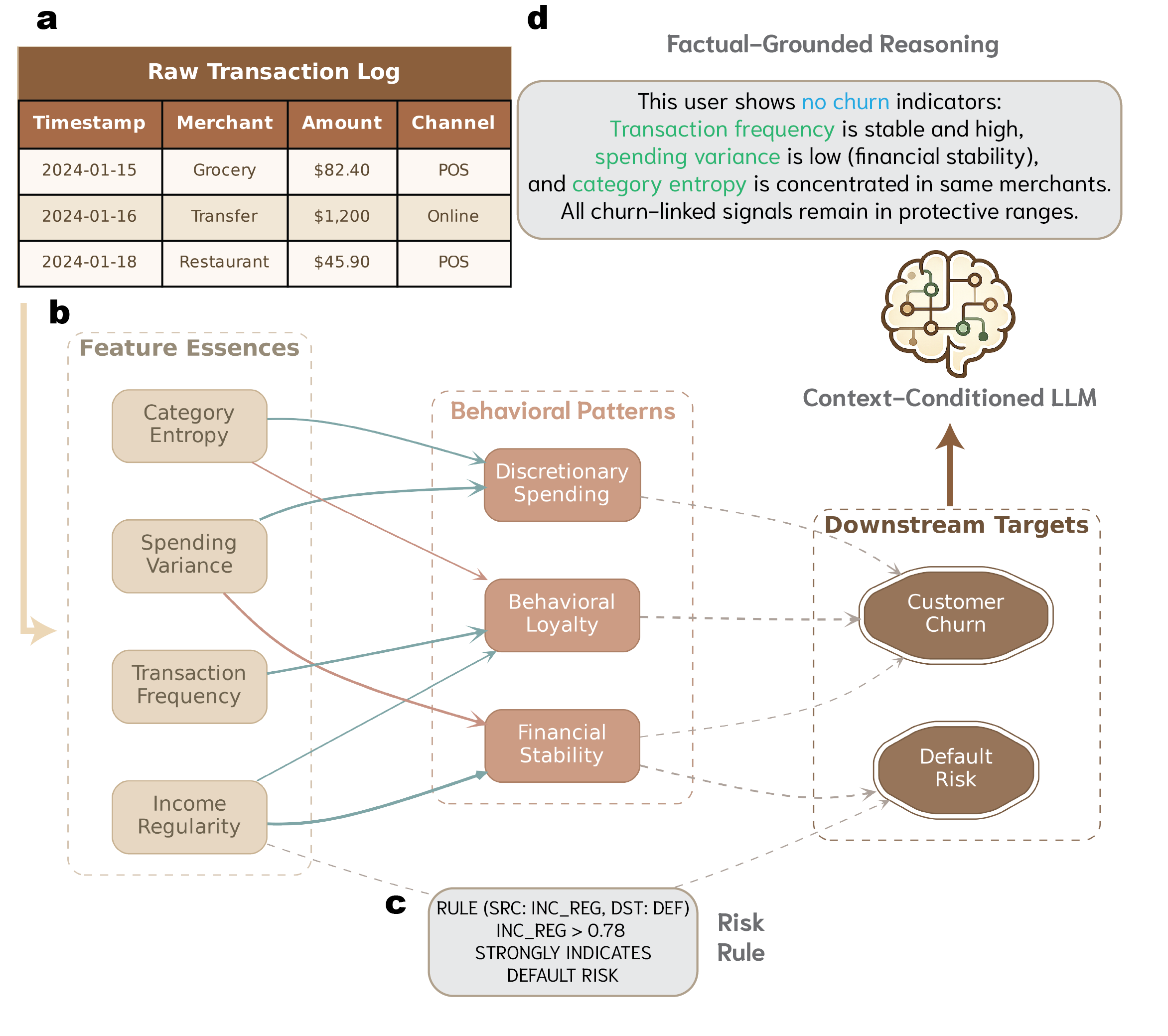}
    \caption{\methodname\ overview. Raw transaction logs (a) are transformed into a structured knowledge base (b) with three semantic layers: feature essences, behavioral patterns, and downstream targets. Explicit white-box rules connect these layers (c), allowing the LLM to produce grounded predictions supported by traceable evidence chains (d).}
    \label{fig:knowledge-graph}
\end{figure}

Figure~\ref{fig:knowledge-graph} shows the overall \methodname\ pipeline. Rather than feeding raw transaction histories (Fig.\ref{fig:knowledge-graph}a) directly to an LLM, \methodname\ converts them into a structured KB that captures reusable behavioral evidence. The KB (Fig.\ref{fig:knowledge-graph}b) is organized into three semantic layers. \textbf{Feature essences} are task-agnostic numerical descriptors computed directly from event sequences, following the transactional feature construction pipeline of LATTE~\cite{latte}. \textbf{Behavioral patterns} are higher-level interpretable features derived from combinations of these essences, capturing concepts such as financial stability or behavioral loyalty.
To construct them, we organize feature essences into three predefined semantic categories aligned with the structure of transactional data: \emph{temporal dynamics} (e.g., activity regularity, inter-transaction intervals), \emph{monetary behavior} (e.g., spending variability, income consistency), and \emph{merchant distribution} (e.g., category concentration, diversity of purchases).
Within each category, we prompt an LLM to propose a small set (typically 1-2 per category) of candidate behavioral patterns that summarize the underlying signals. Section~\ref{sec:ablations} provides a detailed comparison of different selection strategies and their impact on performance. \textbf{Targets} correspond to downstream supervised objectives, such as churn or default risk, when labeled data are available.

To build this KB, \methodname\ first computes feature essences from each user’s transaction history. These essences are then linked to behavioral patterns through explicit rules extracted from a white-box predictor based on AutoWoE~\cite{lama} (Fig.\ref{fig:knowledge-graph}c). In particular, for each behavioral pattern, AutoWoE partitions the selected essences into informative value ranges and assigns each range a signed contribution, which is further converted into human-readable fuzzy rules~\cite{singh2025neuro}. For example, a learned rule can be rendered as:
\begin{quote}
\ttfamily
\footnotesize
IF activity\_period\_days $\leq$ 70.5 $\rightarrow$ strong churn signal.
\end{quote}

When labeled targets are available, the same procedure is applied to learn mappings from feature essences and behavioral patterns to task outcomes, producing rule-level associations that are stored alongside the patterns in the KB.

At inference time, the KB is retrieved to form a structured prompt context, replacing row-wise serialization with a set of behavioral facts and rules (up to 20 facts per prompt).
As predictions are conditioned on these explicit mappings, the LLM generates outputs grounded in traceable evidence chains (Fig.~\ref{fig:knowledge-graph}d), improving interpretability and enabling reuse across downstream financial tasks.

\textit{KB-grounded instruction tuning.} 
Following standard synthetic instruction-tuning pipelines~\cite{selfinstruct,wizardlm}, we convert each KB instance into a grounded reasoning example and use the generated triplets (\{instruction, context, response\}) as a supervision for fine-tuning~\cite{zhou2023lima}. For a given user and target, the prompt includes the target description together with the corresponding KB context: feature essences, inferred behavioral patterns, and explicit rule-based links connecting them to the downstream objective. An LLM is then asked to generate a natural-language explanation and final prediction conditioned only on this structured evidence. This allows the model to internalize KB-grounded financial reasoning patterns without manual annotation.

\textit{KB under limited supervision.} 
\methodname\ also supports adaptation when supervision for the downstream target is limited or unavailable. In the zero-shot regime, if a target is not present during KB construction, we do not fit a target-specific predictor. Instead, we retrieve the subset of behavioral patterns whose natural-language semantics are most relevant to the target description and present their associated white-box rules as structured context to the LLM. The prediction is then produced by inference-time task adaptation: the model conditions on the target description together with retrieved behavioral evidence and composes an output for the unseen objective without parameter updates similar to cross-task in-context transfer approaches~\cite{crosstask}.

In the few-shot regime, we augment the context with up to 16 labeled examples and a small memory of self-reflection pairs, following inference-time refinement ideas from reflection-based reasoning~\cite{reasoningbank}. Each pair stores a prior prediction and its outcome. At inference time, the model first makes an initial prediction and then revises it by comparing its reasoning against these stored successes and failures. This provides a lightweight adaptation mechanism under sparse supervision without updating model parameters.

\section{Experiments}

\subsection{Experimental Setup}

\textit{Datasets.} 
All datasets comprise transactional event sequences. We use three public financial benchmarks together with a proprietary industrial dataset.\\
\textbf{1. Gender Prediction}~\cite{sberbank_gender} is a demographic classification benchmark with 8.4K labeled user sequences defined by relative timestamps, merchant category codes, and transaction types.\\
\textbf{2. Rosbank}~\cite{rosbank1} provides transaction histories for 10K users over a three-month window, with churn labels available for 5K users. \\
\textbf{3. DataFusion}~\cite{DataFusion2024} focuses on churn prediction over a six-month horizon using nine months of prior transaction history. The dataset includes 13M transactions from 96K users (64K labeled), with timestamps, merchant category codes, currency, and transaction amounts available for each record. We use a subsample of 19600 users.\\
\textbf{4. Proprietary Dataset.}
We evaluate the proposed framework on a subsample of 10000 users drawn from a large-scale proprietary banking dataset. The task is to predict a categorical user attribute used for downstream personalization, decision-making, and task routing.

\textit{Implementation Details.}
Zero-shot and few-shot evaluations are performed using gpt-oss-120b~\cite{openai2025gptoss120bgptoss20bmodel}, while Llama-3-8B-Instruct~\cite{grattafiori2024llama} is utilized for the fine-tuning experiments. White-box rules are extracted via AutoWoE~\cite{vakhrushev2021lightautoml}. %\maxim{please add citations here ASAP} 
All computations are conducted on a cluster of four NVIDIA A100 GPUs.

\subsection{Main Results}

\begin{table}[t]
\centering
\small
\caption{\textbf{Few-shot performance comparison across datasets.}
Top-3 methods are highlighted:
\textcolor{black}{\colorbox{bestgold}{1st}},
\textcolor{black}{\colorbox{secondsilver}{2nd}},
\textcolor{black}{\colorbox{thirdbronze}{3rd}}.}
\label{tab:main-results}
\resizebox{\columnwidth}{!}{%
\begin{tabular}{lcc|cc|cc}
\toprule
\multirow{2}{*}{\textbf{Method}} 
& \multicolumn{2}{c|}{\textbf{Rosbank}} 
& \multicolumn{2}{c|}{\textbf{Gender}} 
& \multicolumn{2}{c}{\textbf{DataFusion}} \\
\cmidrule(lr){2-3}
\cmidrule(lr){4-5}
\cmidrule(l){6-7}
& \textbf{F1} & \textbf{MCC}
& \textbf{F1} & \textbf{MCC}
& \textbf{F1} & \textbf{MCC} \\
\midrule

gpt-oss (0 shots)
& 0.55 & 0.19 
& \third{0.60} & \third{0.24}
& \third{0.68} & 0.04 \\

gpt-oss (16 shots)
& 0.58 & 0.25
& 0.59 & 0.22
& \third{0.68} & 0.04 \\

gpt-oss + RT (16 shots)
& \third{0.66} & \third{0.30}
& 0.55 & 0.08
& 0.59 & 0.03 \\

TabPFN\_v2 (16 shots)
& 0.49 & 0.01
& \second{0.61} & \second{0.27}
& \second{0.71} & \second{0.09} \\

TabLLM (16 shots)
& 0.59 & 0.25
& 0.53 & 0.09
& 0.42 & 0.00 \\

FeatLLM (16 shots)
& 0.47 & 0.11
& 0.51 & 0.02
& 0.65 & 0.03 \\

KNN + CoLES (16 shots) 
& 0.47 & 0.06
& 0.58 & 0.16
& 0.61 & 0.01 \\

KNN + LLM4ES (16 shots) 
& 0.45 & 0.06
& 0.51 & 0.02
& -- & -- \\

\specialrule{1.1pt}{0pt}{0pt}

\methodname\ (zero-shot) 
& \second{0.69} & \second{0.38}
& \best{0.63} & \best{0.31}
& 0.65 & \third{0.05} \\

\methodname\ (16 shots)
& \best{0.70} & \best{0.40}
& \third{0.60} & \third{0.24}
& \best{0.77} & \best{0.10} \\

\bottomrule
\end{tabular}%
}
\end{table}

\textit{Performance under limited supervision.} 
We compare \methodname\ against several families of few-shot baselines: prompt-based LLMs (\texttt{gpt-oss} in zero-shot and 16-shot settings, lightweight inference-time adaptation via Reflection Tuning~\cite{reasoningbank}), few-shot tabular learners (TabPFN\_v2~\cite{hollmann2025accurate}, TabLLM~\cite{hegselmann2023tabllm}, FeatLLM~\cite{pmlr-v235-han24f}), %\maxim{please insert citations here}, 
and embedding-based classification pipelines built on pretrained transactional encoders (KNN + CoLES~\cite{babaev2022coles}, KNN + LLM4ES~\cite{shestov2025llm4es}). These baselines cover the main alternatives for limited-supervision learning: direct in-context reasoning, lightweight prompt adaptation, specialized tabular modeling, and nearest-neighbor transfer over learned transaction embeddings.

\begin{figure}[h]
    \centering
    \includegraphics[width=1\linewidth]{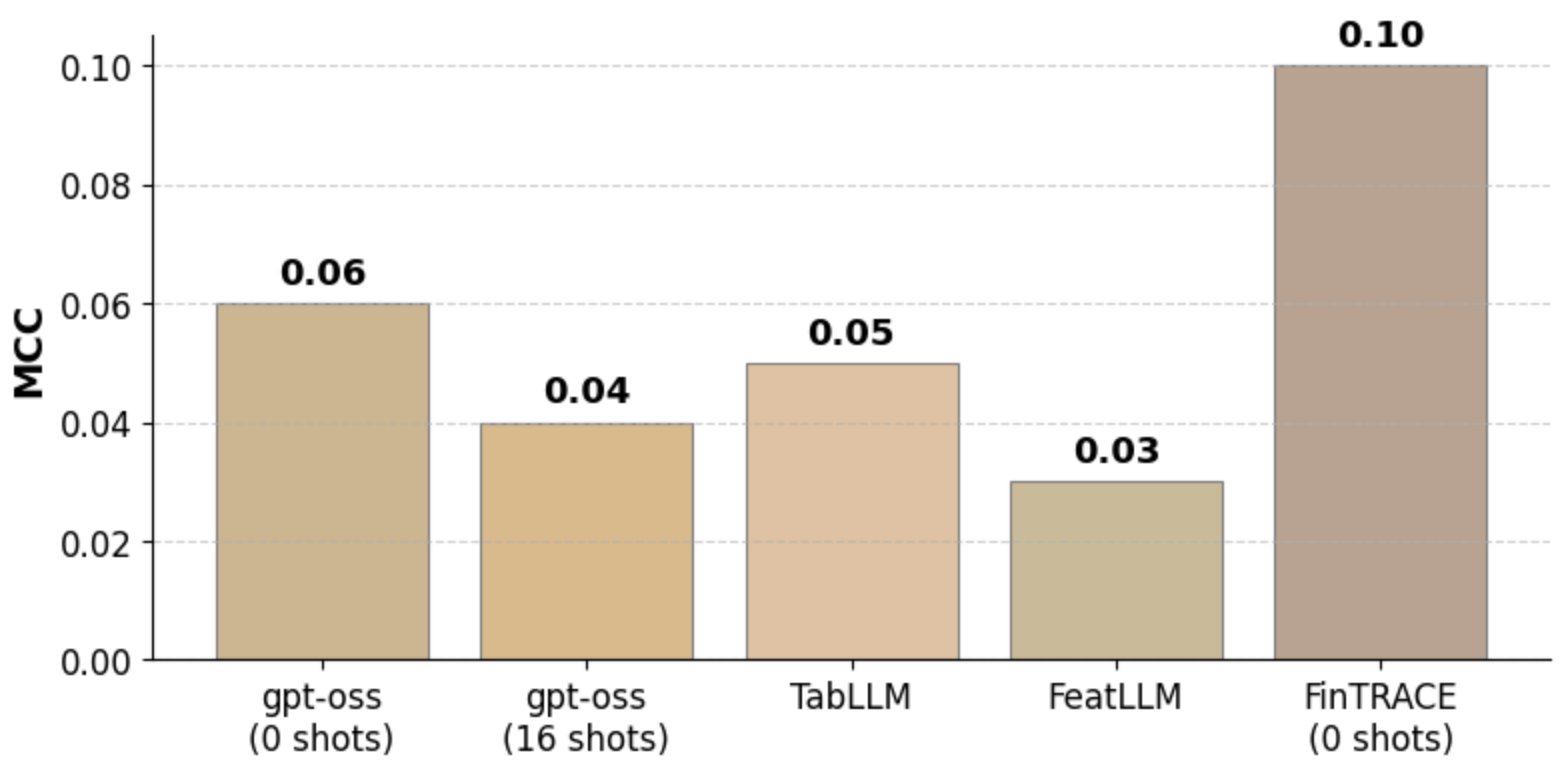}
      \vspace{-8mm}
    \caption{Comparison of LLM approaches on a private dataset.}
    \label{fig:private_dataset}
\end{figure}

Table~\ref{tab:main-results} reports few-shot results on three public benchmarks. FinTRACE achieves the strongest MCC overall while remaining competitive in F1, indicating that behavioral grounding improves label-aware decision quality under scarce supervision. In zero-shot, FinTRACE doubles MCC on Rosbank from 0.19 to 0.38 without task-specific retraining, showing that structured behavioral retrieval provides substantially stronger grounding than prompt-only reasoning. With 16 labeled examples, it reaches 0.40 MCC on Rosbank and 0.77 F1 / 0.10 MCC on DataFusion, outperforming both strong tabular baselines and prior embedding-based transfer methods. 

\textit{Performance on proprietary data.} 
Figure~\ref{fig:private_dataset} extends evaluation to the proprietary dataset under conditions of extreme label scarcity. Consistent with the public benchmarks, zero-shot \methodname\ retains its advantage, achieving the highest MCC (0.10) and outperforming both the base gpt-oss (0.06) and specialized tabular baselines like TabLLM (0.05).

\begin{table}[h]
\centering
\small
\caption{Knowledge-grounded instruction tuning. Relative MCC change is computed with respect to Zero-shot.}
\label{tab:instruction-results}
\begin{tabular}{lcc|cc}
\toprule
\multirow{2}{*}{\textbf{Method}} 
& \multicolumn{2}{c|}{\textbf{Rosbank}} 
& \multicolumn{2}{c}{\textbf{Gender}} \\
\cmidrule(lr){2-3}
\cmidrule(l){4-5}
& MCC $\uparrow$ & Text $\uparrow$
& MCC $\uparrow$ & Text $\uparrow$ \\
\midrule
Zero-shot 
& 0.13 & 0.53 
& 0.18 & 0.53 \\

\methodname\ (target-aware)
& 0.41 {\gain{215\%}} & 0.53
& 0.37 {\gain{105\%}} & 0.53 \\
\midrule

NTP (LLM4ES) 
& 0.40 {\gain{208\%}} & 0.00 
& 0.04 {\loss{78\%}} & 0.00 \\

SFT 
& 0.43 {\gain{231\%}} & 0.53 
& 0.52 {\gain{189\%}} & 0.52 \\

SFT + Instruct 
& 0.03 {\loss{77\%}} & 0.53 
& 0.00 {\loss{100\%}} & 0.52 \\

\methodname\ Instruct 
& \textbf{0.48} {\gain{269\%}} & 0.53 
& \textbf{0.53} {\gain{194\%}} & 0.53 \\
\bottomrule
\end{tabular}
\end{table}

\textit{Knowledge-grounded instruction tuning.} Table~\ref{tab:instruction-results} compares adaptation strategies for transactional prediction—(i) a non-finetuned KB-grounded LLM, (ii) NTP fine-tuning (LLM4ES)~\cite{shestov2025llm4es}, (iii) label-only tuning, and (iv) instruction tuning on knowledge-grounded data. We report MCC on Rosbank and Gender together with general text ability on MMLU~\cite{hendryckstest2021}. \methodname\ Instruct achieves the best trade-off, reaching 0.48 MCC on Rosbank and 0.53 on Gender while preserving text quality. Notably, the Rosbank result matches specialized SOTA transactional models such as CoLES (0.48; Fig.~\ref{fig:label-efficiency}), indicating that knowledge-grounded instruction tuning can close the gap to task-specific architectures without sacrificing general capability. Knowledge grounding alone delivers substantial gains (0.41/0.37 MCC on Rosbank/Gender), bringing the non-finetuned model close to trained adapters. Standard SFT remains competitive, but instruction-style prompting at inference time without instruction-oriented training sharply reduces accuracy; NTP improves Rosbank MCC yet collapses on MMLU, suggesting weaker retention of general language abilities.

% \textit{Knowledge-grounded instruction tuning.} Table~\ref{tab:instruction-results} compares adaptation strategies for transactional prediction, including a non-finetuned KB-grounded LLM, NTP fine-tuning~\cite{shestov2025llm4es}, label-only tuning, and instruction tuning on the knowledge-grounded data. We report downstream MCC on Rosbank and Gender together with preservation of general text ability, measured on MMLU~\cite{hendryckstest2021}. \methodname\ Instruct performs best on both datasets, reaching 0.48 MCC on Rosbank and 0.53 on Gender while maintaining text quality. Notably, the Rosbank result matches the performance of specialized state-of-the-art transactional models such as CoLES (0.48; see Fig.~\ref{fig:label-efficiency}), showing that knowledge-grounded instruction tuning can close the gap to task-specific architectures without sacrificing general language capability. Knowledge grounding by itself delivers a substantial MCC gain (0.41 on Rosbank and 0.37 on Gender), bringing non-finetuned model close to the performance of trained adaptation methods. Standard SFT remains competitive, but simply switching to instruction-style prompting at inference time, without instruction-oriented training, leads to a severe drop in accuracy. NTP (LLM4ES) improves Rosbank MCC but collapses on the text metric, suggesting weaker retention of general language abilities.

\subsection{Ablation studies}
\label{sec:ablations}

\begin{table}[!h]
\centering
\small
\setlength{\tabcolsep}{4pt}      % default is usually 6pt
\renewcommand{\arraystretch}{0.9} % default is 1.0
\caption{Impact of behavioral pattern selection strategy on Rosbank dataset.}
\label{tab:pattern-selection}
\begin{tabular}{lccc}
\toprule
 & \textbf{Random} & \textbf{LLM-guided} & \textbf{Without White-box} \\
\midrule
F1  & 0.63 & 0.69 & 0.41 \\
MCC & 0.28 & 0.38 & 0.01 \\
\bottomrule
\end{tabular}
\end{table}

\textit{Impact of behavioral pattern selection.}
Table~\ref{tab:pattern-selection} compares three strategies for selecting behavioral patterns during KB construction: random selection, LLM-guided selection, and a variant without the White-box component. \methodname\ performs best with LLM-guided selection (F1 \(=0.69\), MCC \(=0.38\)). Random selection is weaker but still substantially outperforms the setup without White-box knowledge, indicating that explicit feature relations are the main source of improvement, while LLM guidance further helps prioritize the most informative patterns.

\begin{table}[!h]
\centering
\small
\caption{Impact of context construction strategy (Rosbank).}
\label{tab:context-ablation}
\begin{tabular}{lccccc}
\toprule
 & \textbf{ZS} & \textbf{+Q} & \textbf{+FI} & \textbf{+Q+FI} & \textbf{KB via WB} \\
\midrule
F1 & 0.55 & 0.54 & 0.51 & 0.55 & \textbf{0.69} \\
MCC                 & 0.19 & 0.17 & 0.10 & 0.18 & \textbf{0.41} \\
\bottomrule
\end{tabular}
\end{table}

\textit{Context construction strategy.} 
Table~\ref{tab:context-ablation} compares different context construction approaches on Rosbank.
\textbf{ZS} denotes vanilla zero-shot prompting over serialized transactions without additional structure.
\textbf{+Q} augments the prompt with simple distributional summaries (feature quantiles), while \textbf{+FI} adds feature-importance scores derived from a lightweight tabular model; \textbf{+Q+FI} combines both heuristics.
These statistical augmentations fail to improve performance, whereas target-aware KB via WB achieves 0.41 MCC, demonstrating that white-box behavioral rules provide the necessary semantic richness for effective inference.

\begin{figure}[!h]
  \centering
  \includegraphics[width=1\columnwidth]{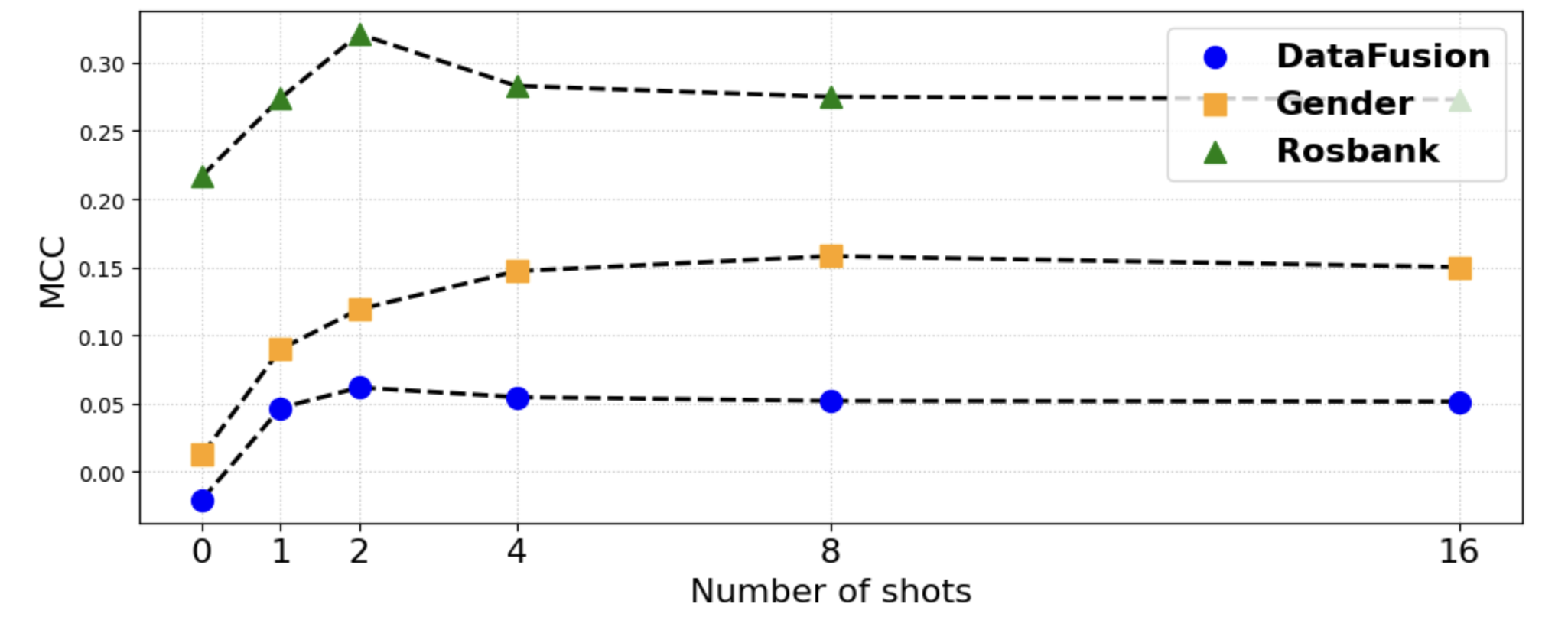}
  \vspace{-8mm}
  \caption{Impact of reflection across different shot budgets.}
  \label{fig:rb-impact}
\end{figure}

\textit{Impact of reflection tuning on few-shot LLM.} 
We isolate the effect of Reflection Tuning (RT) as a lightweight adaptation mechanism applied on top of raw transaction descriptions from LATTE~\cite{latte}. Figure~\ref{fig:rb-impact} reports MCC under different shot budgets, averaged over 5 runs. Across all datasets, reflection improves over the zero-shot baseline, with the largest gains appearing in the first 1-2 shots. Performance then largely saturates, suggesting diminishing returns from additional reflective context. In practice, these results indicate little benefit in going beyond 16 samples.

% --------------------------------------------------

\section{Conclusion}
We presented \methodname, a retrieval-first framework that transforms raw transactional event sequences into a structured behavioral knowledge base for grounded LLM reasoning. Instead of relying on direct row-to-text serialization or task-specific retraining, \methodname\ organizes reusable behavioral patterns and explicit white-box rules, allowing LLMs to reason over interpretable financial evidence. Across public and proprietary benchmarks, \methodname\ substantially improves low-supervision transaction analytics, doubling zero-shot MCC on Rosbank, achieving strong few-shot results across datasets, and enabling knowledge-grounded instruction tuning that achieves state-of-the-art scores for LLMs on open-source financial datasets while preserving general text capabilities.
Overall, these results show that structured behavioral retrieval provides a practical bridge between transactional analytics and language-based reasoning, while opening directions for ontology construction, adaptive memory, and transfer across behavioral abstractions.

\bibliographystyle{ACM-Reference-Format}
\bibliography{sample-sigconf}

\section*{Presenter Biography}
Artem Sakhno is the Research Scientist in the R\&D laboratory at one of the largest banks in Europe (Sber). His work focuses on event sequence modeling, representation learning, sequential user behavior analysis, and the transfer of research advances into large-scale production systems. His research has been presented at multiple top-tier machine learning and NLP conferences, including IJCAI and EMNLP.
\end{document}